\documentclass[useAMS,usenatbib,usegraphicx]{mn2e}
\usepackage{amsmath}
\usepackage{amssymb}
\usepackage{color}
\usepackage{float}

\newif\ifAMStwofonts
\AMStwofontstrue

\newcommand{\simlt}{\lower.5ex\hbox{$\; \buildrel < \over \sim \;$}}
\newcommand{\simgt}{\lower.5ex\hbox{$\; \buildrel > \over \sim \;$}}

\title[A deeper look at the dust attenuation law]
{A deeper look at the dust attenuation law of star-forming galaxies at high redshift}

\author[M. Tress et al.]
{M. Tress$^1$,
 I. Ferreras$^{1,2,3}$\thanks{E-mail: iferreras@iac.es}, P.~G. P\'erez-Gonz\'alez$^{4,5}$, 
 A. Bressan$^6$, \and
 G. Barro$^7$, H. Dom\'\i nguez-S\'anchez$^8$, C. Eliche-Moral$^5$\\
$^1$ Mullard Space Science Laboratory, University College London, 
Holmbury St Mary, Dorking, Surrey RH5 6NT, UK\\
$^2$ Instituto de Astrof{\'i}sica de Canarias, Calle V{\'i}a L{\'a}ctea s/n,
E38205, La Laguna, Tenerife, Spain\\
$^3$ Departamento de Astrof{\'i}sica, Universidad de La Laguna (ULL), E-38206 La Laguna,
Tenerife, Spain\\
$^4$ Centro de Astrobiolog\'{\i}a (CAB, INTA-CSIC), Carretera de Ajalvir km 4, E-28850 Torrej\'on de Ardoz, Madrid, Spain\\
$^5$ Departamento de Astrof\'\i sica, Facultad de CC. F\'\i sicas, Universidad Complutense de Madrid, E-28040 Madrid, Spain\\
$^6$ SISSA, via Bonomea 265, I-34136 Trieste, Italy\\
$^7$ Department of Physics, University of the Pacific, Stockton, CA 95211, USA\\
$^8$ Department of Physics and Astronomy, University of Pennsylvania, Philadelphia, PA 19104, USA\\
}

\begin{document}
\date{MNRAS, Accepted 2019 July 02. Received 2019 July 01; in original form 2018 December 10}
\pagerange{\pageref{firstpage}--\pageref{lastpage}} \pubyear{2019}
\maketitle
\label{firstpage}


\begin{abstract}
A diverse range of dust attenuation laws is found in star-forming
galaxies. In particular, \citet{TMF:18} studied the SHARDS survey to
constrain the NUV bump strength ($B$) and the total-to-selective ratio
($R_{V}$) of 1,753 star-forming galaxies in the GOODS-N field at
1.5$<$z$<$3.  We revisit here this sample to assess the implications
and possible causes of the correlation found between $R_V$ and $B$.
The UVJ bicolour plot and main sequence of
star formation are scrutinised to look for clues into the observed
trend. The standard
boundary between quiescent and star-forming galaxies is preserved when
taking into account the wide range of attenuation parameters. However, an
additional degeneracy -- regarding the effective attenuation law -- is
added to the standard loci of star-forming galaxies in the UVJ diagram.
A simple phenomenological model
with an age-dependent extinction (at fixed dust composition) is
compatible with the observed trend between $R_V$ and $B$, whereby the
opacity decreases with the age of the populations, resulting in a
weaker NUV bump when the overall attenuation is shallower (greyer). In
addition, we compare the constraints obtained by the SHARDS sample
with dust models from the literature, supporting a scenario
where geometry could potentially drive the correlation between $R_{V}$
and $B$.
\end{abstract} 


\section{Introduction}
\label{Sec:Intro}

Despite the fact that dust particles only account for less than 1 per cent
in mass of the interstellar medium \citep{GGJ:18}, dust is a key
component in galaxies. Starlight is affected in various ways depending
on wavelength: it is scattered and absorbed, preferentially at shorter
wavelengths and re-emitted in the infrared. For this reason, accurate
dust corrections are fundamental when deriving stellar ages, stellar masses  and star
formation rates from photometry. By dust extinction, we refer both to
the absorption and scattering of light away from the line of sight.
Dust attenuation includes the contribution of the so-called `star-dust
geometry', where light from sources not along the line of sight may
also scatter into it \citep{Cal:01}. A parameter typically used to
characterize the attenuation law is the total-to-selective ratio
$R_{V} \equiv A_{V}/E(B-V)$, where $E(B-V)\equiv A_{B}-A_{V}$ is the colour excess or
reddening and $A_{X}$  is the extinction through
the $X$ band. A lower value of $R_{V}$ implies a steep (strongly
wavelength-dependent) attenuation law, whereas a higher value of $R_V$
produces a greyer (weakly dependent on wavelength) attenuation. In
addition to the smooth dependence of the attenuation law with
wavelength, a number of resonant features are present, most notably
the 2,175\,\AA\ bump in the NUV spectral window \citep{STP:69}. This
absorption feature is likely explained by small
carbon-rich molecules, such as Polyaromatic Hydrocarbons (PAHs), or
amorphous carbon \citep{GGJ:18}. It is present both in the
extinction law of the Milky Way (MW) and the Large Magellanic Cloud
(LMC), but absent in the metal poorer Small Magellanic Cloud
\citep[SMC,][]{Pei:92},
although \citet{HSH:17} report a milder NUV bump towards the
north-east of the SMC.

Observations of individual stars in our Galaxy towards different
sightlines reveal a diverse range of extinction
laws \citep{CCM:89,FM:07}, with similar results in nearby galaxies
such as the LMC and the SMC \citep{GCM:03, HSH:17, GGJ:18}. For
instance, in our Galaxy, we observe a wide range of $R_{V}$, varying
from 2 to 5, with an average of 3.1, adopted as the standard
extinction law of the MW \citep[see e.g.][]{F:99}.  The
non-universality of the dust extinction law is also seen in nearby
galaxy M31 \citep[]{CGB:15}. \citet{Tang:14,Tang:16} found that the
more massive stars in resolved dwarf irregular galaxies are more
reddened than less massive stars, indicating an age-selective
extinction.  Further afield, variations of the dust attenuation law
(i.e. no longer considering single stars in the analysis) are found in
the starburst galaxy M82, showing a radial gradient towards a steeper
law and a weaker bump at large galactocentric
distance \citep[]{HF:15}.
\citet[]{Cal:94,Cal:00} provide a comprehensive analysis of
nearby starburst galaxies. The retrieved average attenuation curve is called
the Calzetti law, frequently adopted to describe the effect of dust in 
starburst galaxies at all redshifts, and often applied to non starburst galaxies as well.
This attenuation law lacks a bump and it is greyer (R$_V$=4.05) than that of
the MW. \citet{CSB:10} constrained the dust attenuation in a sample of
low redshift galaxies combining {\sl GALEX} and SDSS photometry,
and found an overall weaker NUV bump with
respect to our Galaxy. \citet{WCB:11}
explored a large sample (23,000) of star-forming galaxies, finding
dust attenuation curves that were, on average, consistent with the presence of
an NUV bump, with variations most likely influenced by the dust geometry.  In
contrast, the study of \citet{BCC:16, BCC:17} found no confirmation of the NUV bump,
with shallow attenuation curves. More recently, \citet[]{SBL:18} studied
the dust attenuation law of more than 200,000 galaxies at low 
redshift. They observed a wide range of NUV bump strengths and
dust attenuation slopes. In this study, steeper curves were found to 
correlate with the optical opacity, specifically higher opacities 
leading to a flatter curve.

Comparable results are obtained at higher redshift.  For
instance, \citet[]{BBI:05} conclude that their sample of UV-selected
galaxies have dust attenuation curves similar to the LMC extinction
law, but FIR-selected galaxies are similar to the MW
curve. \citet[]{NPS:07} studied a sample of 108 massive star-forming
galaxies at 1$<$z$<$2.5. The NUV bump was present in a third of their
sample. Galaxies with redder UV SEDs were associated with the presence
of the carriers of the NUV bump. The study of z$\sim$2 galaxies of 
\citet[]{BNB:12} found that the UV dust attenuation increases
with stellar mass and drops when the UV luminosity increases. Moreover, 
they detected the NUV bump in 20\% of the galaxies.
Additionally, \citet[]{KC:13} used composite SEDs to explore the dust
attenuation curve in a set of galaxies at 0.5$<$z$<$2.0.  They found
that the best-fit slope and the NUV bump strength are correlated, with
steeper laws associated with a stronger bump.  The \citet[]{RKS:15}
study on galaxies at z$\sim$2 found that, at shorter wavelengths, the
dust curve has a similar shape as the Calzetti law, comparable to the
SMC at $\lambda\gtrsim$2,500\,\AA. The study of 
\citet[]{SPL:16} with galaxies at redshift z$\sim$1.5-3 found that a lower (higher) colour
excess corresponds to a steeper (flatter) dust curve. However, they
observe no correlation with other galaxy properties.

In general, studies provide evidence supporting the presence of a
diverse range of attenuation
properties \citep[see,e.g.][]{BCC:17,JSS:07,WCB:11,BBI:05,BNB:12,CSB:10}. In
line with these studies, an analysis constraining the dust attenuation
law at higher redshift was recently presented in \citet[]{TMF:18}. The medium band
(R$\equiv \lambda/\Delta\lambda\sim 50$), deep ($\simlt$26.5\,AB) optical photo-spectra of the
Survey of High-z Absorption Red and Dead Sources \citep[SHARDS,][]{SHARDS} provides an
optimal dataset to probe the dust attenuation law in the rest-frame
NUV region of galaxies at redshift z$\sim$2. The results from this  work agree
qualitatively with \citet[]{KC:13} and \citet[]{SBL:18}, specifically
a strong correlation is found between $R_{V}$ and $B$, whereby a steeper attenuation
curve is associated with a stronger bump. A decreasing
colour excess with increasing bump strength was also found.

Theoretical dust models allow us to account for the diversity of the
dust attenuation laws. \citet[]{CF:00} strive to depict different
effects of dust and its geometry. In this case, the younger population
suffers not only the attenuation due to the presence of their own
birth clouds but also the diffuse interstellar medium (ISM). Older stars, in turn, are
affected only by the latter \citep[e.g.][]{PGB:07}. The study favours
a mixed-slab model for the ISM. Along the same lines, \citet[]{WG:00}
use multiple-scattering radiative transfer calculations to represent
different types of galactic environment and dust composition (MW and
SMC). Flatter dust curves with a weaker bump are found in clumpier
dust distributions. \citet[]{DL:07} calculated the IR
emission using a silicate-graphite-PAH mixture dust model that 
agree with the extinction in the MW. More
recently, \citet[]{SD:16} produced a radiative transfer model in a
spherical, clumpy interstellar medium (ISM). Radiative transfer
effects cause weaker NUV bumps along with greyer attenuation curves
when the ISM is clumpier and dustier.  Additionally, a physical model
for the presence of these variations in the dust attenuation law in
galaxies was recently presented by \citet[]{NCDJP:18}. They study cosmological
hydrodynamic simulations of galaxy formation, implementing a 3D Monte
Carlo dust radiative transfer code. They found that, even
after assuming the same underlying dust composition, the attenuation
laws rendered a wide range of properties. These variations are
dependent on a diverse star-dust geometry. For instance, the different
values of the NUV bump strength is associated to the fraction of
unobscured O and B stars in their model.

The goal of this paper is to extend the analysis of the observed
constraints on the dust attenuation law presented in \citet[]{TMF:18},
specifically to study the correlation between $R_{V}$ and $B$, to shed
light on the degeneracy between dust composition and the star-dust
geometry, as potential drivers of this trend.  The structure of the paper is as
follows. In \S\ref{Sec:UVJMS} we reassess the SHARDS sample
from \citet[]{TMF:18}, exploiting its position on the UVJ bicolour
diagram and on the main sequence of star formation in galaxies.
\S\ref{Sec:mutau} presents a simple phenomenological model based on 
an age-dependent attenuation (at constant extinction, i.e. dust
composition) that produces an effective attenuation law with similar
properties to those observed in the SHARDS
sample. \S\ref{Sec:Modelseq} is devoted to a comparison of the
observed trends with two dust models, namely \citet[]{DL:07}
and \citet{WG:00}. We conclude in \S\ref{Sec:Disc} with a discussion 
that wraps together the results obtained from the different tests performed in
this paper.

\section{Revisiting the SHARDS star-forming galaxy sample}
\label{Sec:UVJMS}

This paper is motivated by a study of dust attenuation in a sample of
star-forming galaxies compiled from SHARDS, the Survey for High-z
Absorption Red and Dead Sources \citep{SHARDS}.  This study was
presented in \citet[]{TMF:18}.  The SHARDS survey is an ultra deep
galaxy survey towards the GOODS North field, comprising medium-band
photometry (FWHM$\sim$150\AA) in the optical window (5,000--9,500\AA).
The observed photometry in all 25 medium-band SHARDS passbands is
combined with PSF-matched broadband photometry from HST in the optical
and NIR (F435W, F105W, F125W, F140W, F160W) as well as Spitzer/IRAC
3.6\,$\mu$m fluxes and ground-based K$_s$ data (see fig.~2
of \citealt{TMF:18}). The sample comprises 1,753 galaxies with high
enough SNR ($\geq$5) across all SHARDS passbands, and cover
the redshift window 1.5$<$z$<$3. The sample includes estimates of
stellar mass, star formation rate and average, SSP-equivalent, stellar
ages. The fluxes are compared with population synthesis models over
a wide range of ages and chemical composition, and include
dust attenuation as a foreground screen, following
the generic law proposed by \citet[][hereafter
CSB10]{CSB:10}. This law is fully defined by three free parameters:
the total to selective extinction ($R_V$), the NUV bump strength
($B$), and the colour excess, $E(B-V)$. We note that the
alternative parameterisation proposed by \citet{Noll:09}
produces very similar results in our analysis. Appendix~A of 
\citet{TMF:18} provides a simple functional mapping between these
two choices. We follow a standard Bayesian approach, with flat
priors on the parameters, to derive the a posteriori probability
distribution function of the parameters, from which the best estimates
and uncertainties are derived. See \cite{TMF:18} for more details
about the fitting procedure, including comparisons with mock data
following a wide range of star formation histories.

Our analysis of SHARDS star-forming galaxies revealed a wide range of
attenuation laws. Most importantly, the dust-related parameters
feature a significant correlation between $R_{V}$ and $B$: as the dust
attenuation law becomes flatter (higher $R_V$), the NUV bump
weakens. This trend can have two possible causes. Firstly, as a result
of a variation in the grain size distribution, the observed trend
would imply that smaller dust grains are associated with a stronger
NUV bump. A second explanation to this variation involves changes in
the dust geometry among galaxies, i.e. the distribution of dust with
respect to the underlying stellar populations.

This section extends the analysis of  \citet{TMF:18}, 
studying the distribution on the rest-frame UVJ colour-colour diagram, and
seeking potential trends of the parameters with respect to the main sequence
of star-forming galaxies.

\subsection{UVJ diagram}
\label{UVJ}

We now turn our attention to the rest-frame UVJ colour-colour
diagram. This diagram combines two different spectral intervals, and
provides a relatively ``cheap'' diagnostic, in terms of telescope
time, to disentangle the effects of dust from age (or metallicity).
This technique is relatively widespread in the analysis of deep
photometric data. A galaxy can appear red because the underlying
stellar populations are old, therefore dominated by cool, evolved
low-mass stars.  However, a young, dust enshrouded galaxy could appear
equally red, giving rise to the dust-age degeneracy. By contrasting
the colour from a short interval in the region of the age-sensitive
4,000\,\AA\ break ($U-V$) with a wider one extending from the optical
to the near infrared ($V-J$), it is possible to discriminate between
an old-dustless and a young-dusty galaxy. This diagram has been
profusely used in the literature to classify galaxies into quiescent
and star-forming
systems \citep[e.g.][]{Labbe:05,Williams:09,Whitaker:11,Cava:15,HDS:16,LADG:17}.

\begin{figure*}
\centering
\includegraphics[width=170mm]{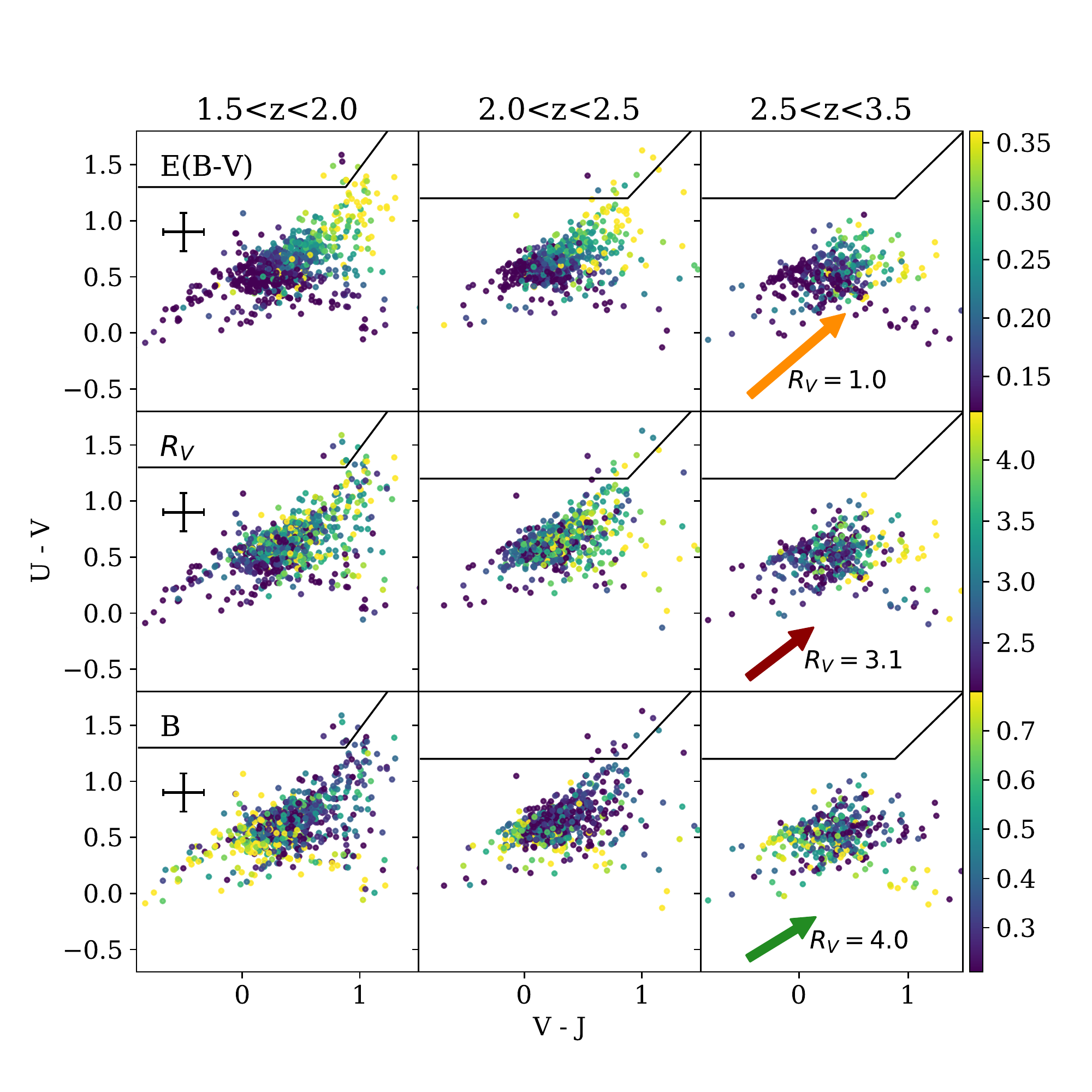}
\caption{Rest-frame UVJ bicolour diagram of the SHARDS star forming
sample from \citet{TMF:18}. The distribution is colour coded with respect to
the best-fit dust parameters, as labelled (from top to bottom,
$E(B-V)$, $R_{V}$ and $B$, with colour bars shown to the right in each row).
The sample is shown in three bins
of redshift (increasing from left to right), and we include the boundary 
between quiescent and star-forming galaxies, as 
defined in \citet{Whitaker:11}. The arrows on the rightmost panels give the
dust correction at $A_{V} = 1$ at three different values of
$R_{V}= \{1.0, 3.1, 4.0\}$, as labelled.}
\label{fig:uvjdust}
\end{figure*}

However, the large variations found in the dust attenuation parameters
could, in principle, affect the photometric data in ways that could
alter the border between quiescent and star-forming galaxies.
After all, the dust correction vector could shift and ``put''
some of the dusty star forming galaxies in the region of the UVJ diagram traditionally
classified as quiescent -- when adopting a fixed dust attenuation law.
We studied this point by identifying our SHARDS
sample on the UVJ plane (Fig.~\ref{fig:uvjdust}).  We obtain the
rest-frame $U-V$ and $V-J$ colours by use of the stellar population
synthesis models of \citet[][hereafter BC03]{BC03}, adopting the
best-fit stellar population and dust-related parameters. These
best-fit values are obtained via a comparison of the SHARDS photometry
as well as several NIR broadband fluxes with the BC03 models,
including the generic dust extinction model of CSB10 \citep[see][for
details]{TMF:18}.  We note the sample was selected by imposing a
threshold in the SNR ($\geq$5) in {\sl all} the SHARDS (optical) passbands
of galaxies over
the redshift range 1.5$<$z$<$3. This selection implies the galaxies
have a high enough flux in the rest-frame NUV, meaning the
sample should be made up of star-forming systems.  As expected, the
majority of our galaxies live on the star-forming region of the UVJ
diagram, extending over a wide diagonal strip running from the
bottom-left to the top-right, that mainly corresponds to the effect of
dust attenuation on the colours.

The UVJ diagram in Fig.~\ref{fig:uvjdust} is shown in several panels,
dividing the sample with respect to redshift (left to right in
increasing order of z), colour coding the symbols with respect to each
of the three dust-related parameters -- from top to bottom: $E(B-V)$;
$R_V$ and $B$.  A typical error bar is shown on the leftmost panels
(at the 1\,$\sigma$ level).  The solid line limits the region of
quiescent galaxies (to the top-left of the boundary), following the
redshift-dependent definition of \citet{Whitaker:11}.  In general, the
presence of the bump in a galaxy does not seem to largely affect its position
within the UVJ diagram, although there is a preference towards
stronger bumps in galaxies towards the blue-blue (i.e. bottom-left)
part of the diagram. The total-to-selective ratio also
shows a trend, with steeper (i.e. lower R$_V$) attenuation towards
the same part of the diagram. As expected, the colour excess, $E(B-V)$,
shows the strongest trend, with higher values towards the top-right
part of the diagram. To quantify these trends, we calculated the
Kolmogorov-Smirnov (KS) statistic, extended to two-dimensional datasets,
as implemented by \citet{Peacock:83} and \citet{FF:87}.
In each case, we split the sample at the median value of the given
dust parameter, and compare the resulting samples with the KS test. We
find a KS statistic $D_{\rm KS}$ of 0.54 (when splitting the sample
with respect to $E(B-V)$); 0.32 (for a $R_V$ split); 0.30 (for a $B$
split).  To assess the significance of these values, we performed 1,000
random splits of the sample (i.e. creating subsets with the same
number of galaxies) producing a Monte Carlo
distribution that gave $D_{\rm KS}=0.06\pm 0.01$ (1\,$\sigma$), confirming that the
trends with respect to the dust related parameters are statistically
significant, with the most significant one being the split in colour
excess.

At the highest redshift bin (rightmost panels),
we observe comparatively fewer red galaxies. This may be explained by
an observational bias imposed by the flux and SNR limit of the sample
selection.  The rightmost panel of Fig.~\ref{fig:uvjdust} illustrates
the effect of a varying dust attenuation law by showing three dust
vectors for $A_V=1$, corresponding to, from top to bottom: $R_{V} =\{
1.0, 3.1, 4.0\}$, as labelled, assuming a Milky Way NUV bump strength
($B=1$). Note how a steeper dust attenuation law (lower $R_V$) results
in a higher reddening at fixed $A_{V}$ -- as expected from the
definition of colour excess: $E(B-V)\equiv A_V/R_V$. This figure
confirms that the wide range of dust attenuation values does not
affect the separation between quiescent and star-forming galaxies, as
those variations follow the same direction as changes in dust.
However, {\sl a new degeneracy} is added in the analysis, so that the
diagonal strip of star-forming galaxies span a range of age,
metallicity, colour excess {\sl and} $R_V$.

\begin{figure}
\centering
\includegraphics[width=8.9cm]{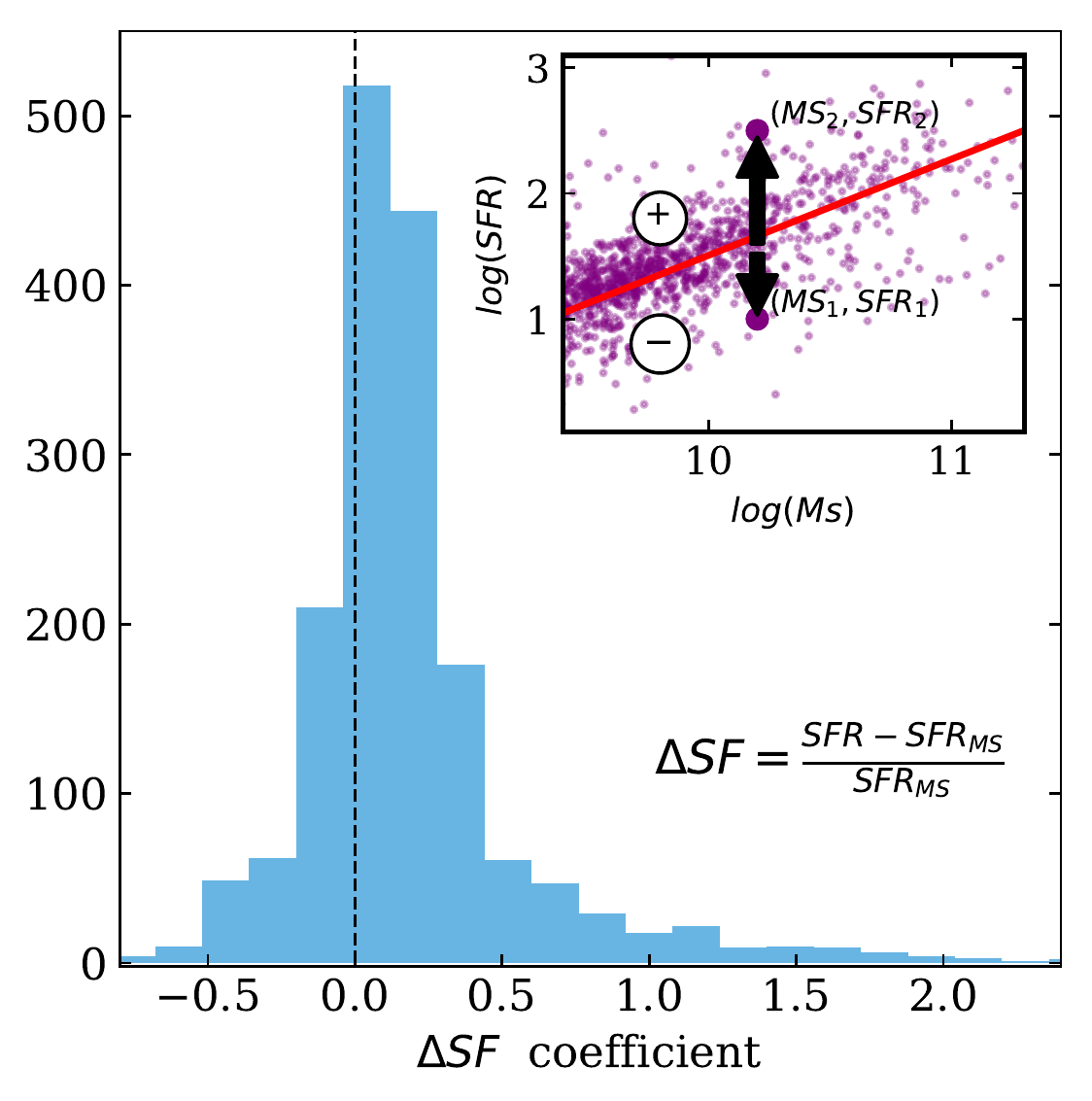}
\caption{Distribution of the $\Delta SF$ parameter of our SHARDS sample.
The definition of this parameter is illustrated in the inset.
$\Delta SF$ has a positive or negative value depending on the
position of the data point with respect to the Main Sequence (MS)
defined on the plane spanned by  $\log$\,SFR vs $\log M_s$.}
\label{fig:MShist}
\end{figure}

\subsection{Main sequence (MS) of star formation}
\label{MS}

In order to characterise further our galaxy sample, we explore the
correlation between the derived star formation rate (SFR) and stellar mass. In
particular, we want to compare the activity with respect
to the Main Sequence (MS) of star forming galaxies. We adopt the
definition of the MS given by \citet{Speagle:14}. It is important to
note that the sample used for this particular exercise corresponds to
objects with well-defined SFRs (i.e. those galaxies that only have 
upper limits are excluded here). In order to parameterise the
location of a galaxy on the SFR vs mass diagram with respect to the
MS, we define a dimensionless coefficient $\Delta SF$. This coefficient represents
the fractional offset in the SFR 
between the targeted galaxy and the Main Sequence, at fixed mass
and at the redshift of the galaxy, namely:
\begin{equation}
\Delta SF(z) =  \frac{SFR - SFR_{MSz}}{SFR_{MSz}},
\end{equation}
where, $SFR_{MSz}$ is the star formation rate at the position of the
MS for the stellar mass and redshift as the targeted galaxy.  The star
formation rates of our sample are taken from \citet{Wuyts:11}, who
use photo-spectra covering a wide range from the $U$ band to 8$\mu$m fluxes.
We use their SFR$_{\rm UV+IR}$ indicator, and note that all our SHARDS sources
have star formation rate estimates. As a
check on the effect of redshift on the derivation of $\Delta SF$,
we compare the analysis using the MS
at the median redshift z$=$2, and then recalculated $\Delta SF$
taking the redshift-dependent MS. The differences between these two methods are negligible.
Fig.~\ref{fig:MShist} shows the distribution of the $\Delta SF$
coefficient.  A positive (negative) value means the observed SFR is
higher (lower) than the one the galaxy should have if it were on the
Main Sequence.  Most of our sample falls close to the MS relation,
with variations mostly within $\pm$35\% (over 77\% of the
sample has $|\Delta{\rm SF}|\leq0.35$).  The inset of Fig.~\ref{fig:MShist} 
illustrates the definition of this coefficient.
We refer the reader to fig.~1 of \citet{TMF:18} for a
standard representation of our SHARDS sample with respect to the MS.

\begin{figure}
\centering
\includegraphics[width=85mm]{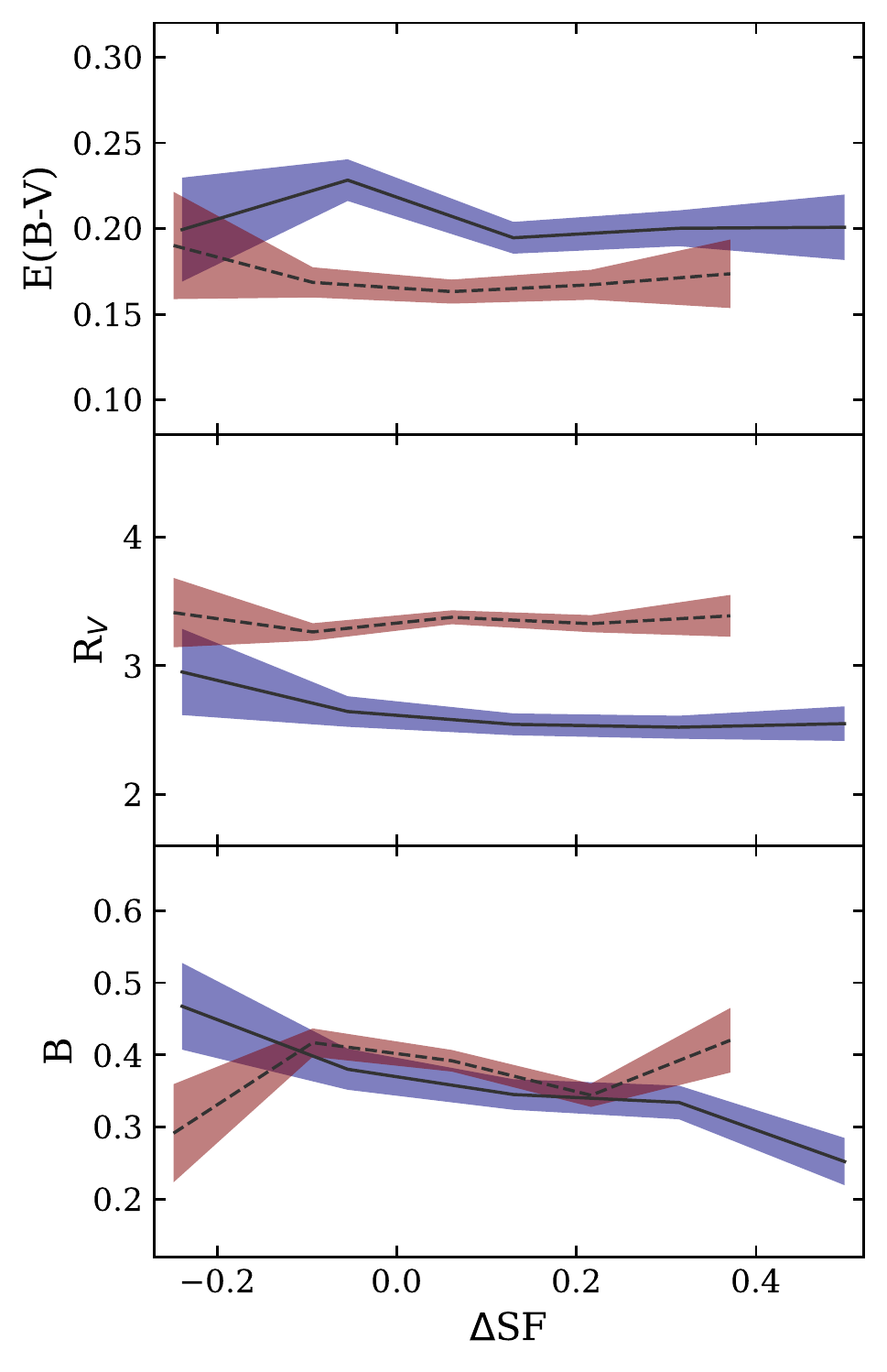}
\caption{Comparison of the $\Delta$SF coefficient
(see Fig.~\ref{fig:MShist}) with the best-fit dust-related
parameters. In all panels, the lines (and shade) trace the median (and
1\,$\sigma$ scatter) of subsamples, split with respect to stellar age
(blue-solid: young tercile, red-dashed: old tercile). The median
stellar age of the distribution is 5.2\,Myr.}
\label{fig:MSdust}
\end{figure}

Fig.~\ref{fig:MSdust} shows the distribution of $\Delta SF$ with
respect to the three dust-related parameters (from top to bottom)
$E(B-V)$, $R_V$ and $B$.  In all panels, the lines (and shade) trace
the median (and 1\,$\sigma$ scatter) of subsamples, split with respect
to stellar age (blue-solid: young tercile, red-dashed: old
tercile). The median stellar age of the distribution is 5.2\,Myr (note,
for reference, the plots regarding the distribution of other
observables of the sample in fig.~7 of \citealt{TMF:18}). Our sample
is slightly biased towards stronger star formation than MS galaxies,
with a median of $\Delta$SF$_{m}=+0.10$, with the offset dominated by
younger galaxies (the median for this subsample, represented by the
blue shaded regions in Fig.~\ref{fig:MSdust} is $+0.17$). Consistently 
with the findings in \citet{TMF:18}, we recover the trend towards higher values of
$R_V$ for the older population, and a higher colour excess in younger star-forming
galaxies. Moreover, the figure suggests a weak decrease of the NUV bump with increasing
$\Delta$SF in the young subsample. However, given the typical error bars expected in the
derivation of $B$ (around 0.2, see table~2 of \citealt{TMF:18}), we
can only present this as a marginal trend.

\section{Age-dependent extinction: a phenomenological model}
\label{Sec:mutau}

The standard approach to study the attenuation law and its relation to the
dust composition and its distribution among the stellar populations
involves radiative transfer modelling (see, e.g., \citealt{WG:00};
\citealt{Groves:04}; \citealt{PGB:07}; \citealt{Popescu:11};
\citealt{Natale:15}; \citealt{SD:16}).  However, a simple
phenomenological model can be explored, that encapsulates the
fundamental aspect dealing with the effect of the distribution of dust
within the stellar populations. We present such model here. We
make the working hypothesis that the extinction law (i.e. the
wavelength dependence of dust scattering and absorption, dependent on
the dust composition) is the same in all galaxies, whereas
{\sl the observed variations in the attenuation law among galaxies are
produced as a result of a different ``geometry''}. 
Dust is expected to mostly affect the youngest populations, typically
found in gas-rich and dust-rich regions. As age progresses, these
populations -- along with the gas and dust content -- disperse,
resulting in weaker attenuation. \citet{Silva:98}
presented for the first time a galaxy model with an age-selective and
geometry dependent extinction. Similarly, \citet{CF:00} proposed a
simple description of attenuation in galaxies involving two different
dust laws: a diffuse one for old populations and a higher opacity
relation for the young components that are still
embedded in very dusty environments.

We extend this model here, adding a smooth time dependence, in
the sense that the overall level of extinction -- parameterised 
by the colour excess, $E(B-V)$ -- for a given, single-age, population
decreases monotonically with time. We assume a constant star
formation rate for this model. The extinction for each component
(and, thus, the dust grain composition) is kept 
fixed at the standard Milky Way law (i.e. $R_V=3.1$ and $B=1$,
adopting the functional form of CSB10). Two functions are considered
for the time dependence: a linear function with (log) time, or an
exponentially decreasing function. In either case, a single parameter
quantifies the timescale over which the effect of dust decreases.  We
emphasize that this is a strong assumption meant to assess whether an
average composition of Milky Way-like dust could lead to 
attenuation trends qualitatively similar to the observations. We also
note that the timescales considered here are significantly longer
than those adopted in the birth cloud model of \citet{CF:00}.
Our model does not aim to follow the physical processes in single
star-forming clouds -- within which dust dispersion and destruction
operate over significantly shorter timescales -- but, rather, to
define a phenomenological parameter that takes into account the global
evolution of the dust content over galaxy scales. Therefore, this
timescale includes the complex aspects of the star formation history
and the evolution of dust geometry. Detailed radiative transfer
models, beyond the scope of this paper, would be required to relate
our longer timescales with the shorter ones typically used in the
birth cloud model.

\subsection{Linear decrease of reddening}
\label{mu}

We build a set of simulated data by combining the BC03 stellar
population synthesis models and adopting an age-dependent
dust attenuation.  We enforce the same distribution of
redshifts, flux uncertainties and SHARDS passband offsets as the original SHARDS
sample presented in \citet{TMF:18}.  We impose the following 
dependence for the colour excess:
\begin{equation}
E(B-V)(t) = \max\left[\epsilon_M -\mu \log\left(\frac{t}{0.01\,{\rm Gyr}}\right), 0\right],
\label{eq:mumodel}
\end{equation}
where the composite stellar population follows a constant star
formation rate between 0.01 and 1\,Gyr. The normalization of this
expression is given by $\epsilon_M$, i.e. the maximum reddening
suffered by the population.  In the simulated data, $\epsilon_M$ is
extracted from a uniform random deviate between 0.1 and 0.6, and $\mu$
is chosen so that the amount of reddening at t=1\,Gyr is zero.
Following this model, a higher $\mu$ produces a rapidly changing
extinction, as if the dust were dispersed very efficiently.
Although this is admittedly a very simple phenomenological
model, our aim is to assess whether an age-dependent dust extinction
model could lead to the correlation found between the dust-related
parameters.

\begin{figure}
\centering
\includegraphics[width=88mm]{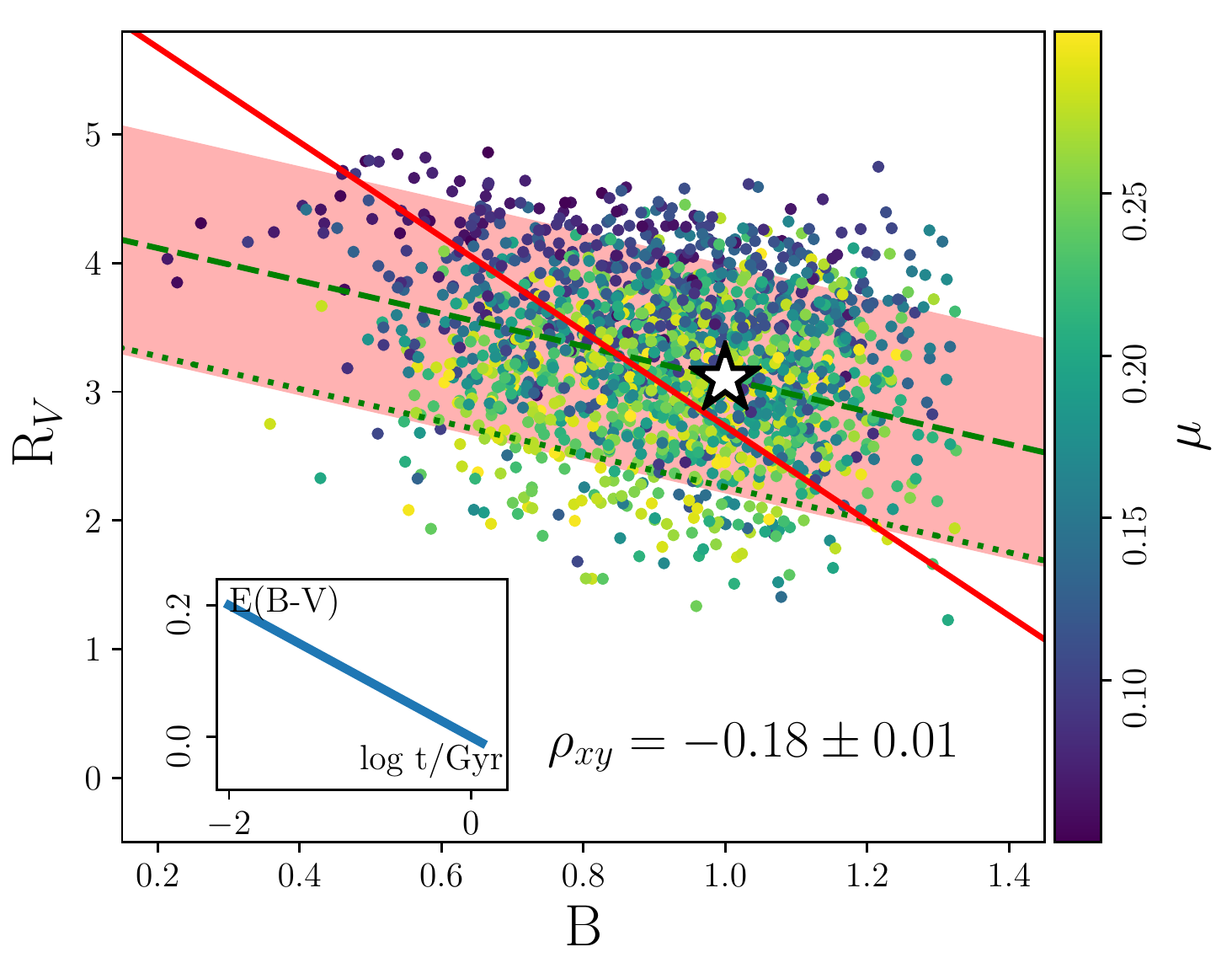}
\caption{Relationship between $R_{V}$ and $B$  when modelling
an age-dependent attenuation that varies linearly with (log) time,
parameterised by a slope $\mu$ (equation~\ref{eq:mumodel}, see inset
for an example).  The dots are colour coded according
to the value of $\mu$ (colour bar on the right). The red solid line
shows the best-fit, and the linear correlation
coefficient ($\rho_{xy}$) is included (along with a 1\,$\sigma$ error bar). The star
symbol marks the standard extinction law of the Milky Way. The dashed
line and shaded area show the observed slope and scatter from \citet{TMF:18}
when imposing the Milky Way constraint at $B=1$, and the dotted line is the
original relation.}
\label{fig:mu}
\end{figure}

Once the mock fluxes are produced, we subject the data to the same
analysis pipeline as the one presented in \citet{TMF:18}, using
exactly the same SSP model grids. The results are shown in
Fig.~\ref{fig:mu}, presenting the trends on the $R_V$ vs $B$
plane. The results are colour coded according to $\mu$. It is
important to note that, by construction, all data feature the same
{\sl intrinsic} (Milky Way) extinction, represented by a star symbol
in the plot. The figure shows that the introduction of an
age-dependent extinction -- acting as a proxy for the dust geometry
-- induces a correlation between the dust parameters of the effective
attenuation law.  The resulting trend (the solid red line is the
result of a linear regression to the data)
agrees {\sl qualitatively} with the observational result 
from \citet[]{TMF:18}, i.e. a stronger bump is expected when the
attenuation law is steeper (lower R$_V$). The dotted line shows
the {\sl observed} trend from that work. A systematic can be expected from a
range of factors, including variations in dust composition, or a more
detailed description of the age-dependent attenuation. To aid the eye,
we shifted the observed trend, imposing the constraint from the Milky
Way extinction curve, so that $R_V=3.1$ at $B=1$, shown as a dashed line,
with a shaded area representing the observed scatter in the trend.

Galaxies with a lower value of $\mu$ (slower time variation) tend to
have a greyer attenuation curve. Galaxies with a
stronger NUV bump have an attenuation curve comparable to the Milky
Way, although there is a wide scatter. We note that the parameter 
uncertainty expected from this methodology is $\Delta R_V\sim 0.7$ and 
$\Delta B\sim 0.2$ (see Tab.~2 in \citealt{TMF:18}).
To quantify the trend introduced by the time variation, we
computed the linear correlation coefficient
between B and R$_V$, obtaining\footnote{We follow the standard
notation, $\rho_{xy}$ for the linear correlation coefficient. In this
case x=$B$ and y=R$_V$.}
$\rho_{xy}= -0.18\pm 0.01$. The error bar (1\,$\sigma$) is obtained from a
Monte Carlo analysis comprising 100 realizations produced by adding
noise to all data points corresponding to the error bars of the
individual parameter estimates. Therefore, the trend is robust. We emphasize that this
exercise is not meant to fully explain the observational trends,
but to motivate how a fixed extinction law could
produce the observed correlations if the dust were to affect the
stellar populations in a time-dependent way.
Variations in the intrinsic dust extinction law -- caused by changes
in the dust composition -- will potentially induce changes that can
explain the mismatch between this phenomenological model and the
observational trends.

This model is qualitatively similar to the `birth model'
of \citet{CF:00}, with younger stars being more attenuated due to the
presence of their birth clouds, and later populations being
progressively less attenuated as the dust is being dispersed or
destroyed. However, we note that the timescales involved in our
model are substantially longer: the classical model
assumes that after $\sim$10\,Myr, the stars break out of the
birth cloud. However, we note that our phenomenological timescale
also incorporates the effect of 
additional processes related to the evolution of dust over galaxy
scales. Regarding the variation of the $B$ parameter, \citet{PGB:07}  predict that
young obscured stars have a weaker bump. Following this exercise,
since the underlying extinction remains the same, we may argue that
geometry plays a fundamental role in the observed trends presented
in \citet{TMF:18}.

\subsection{Exponential decrease of reddening}
\label{tau}

\begin{figure}
\centering
\includegraphics[width=88mm]{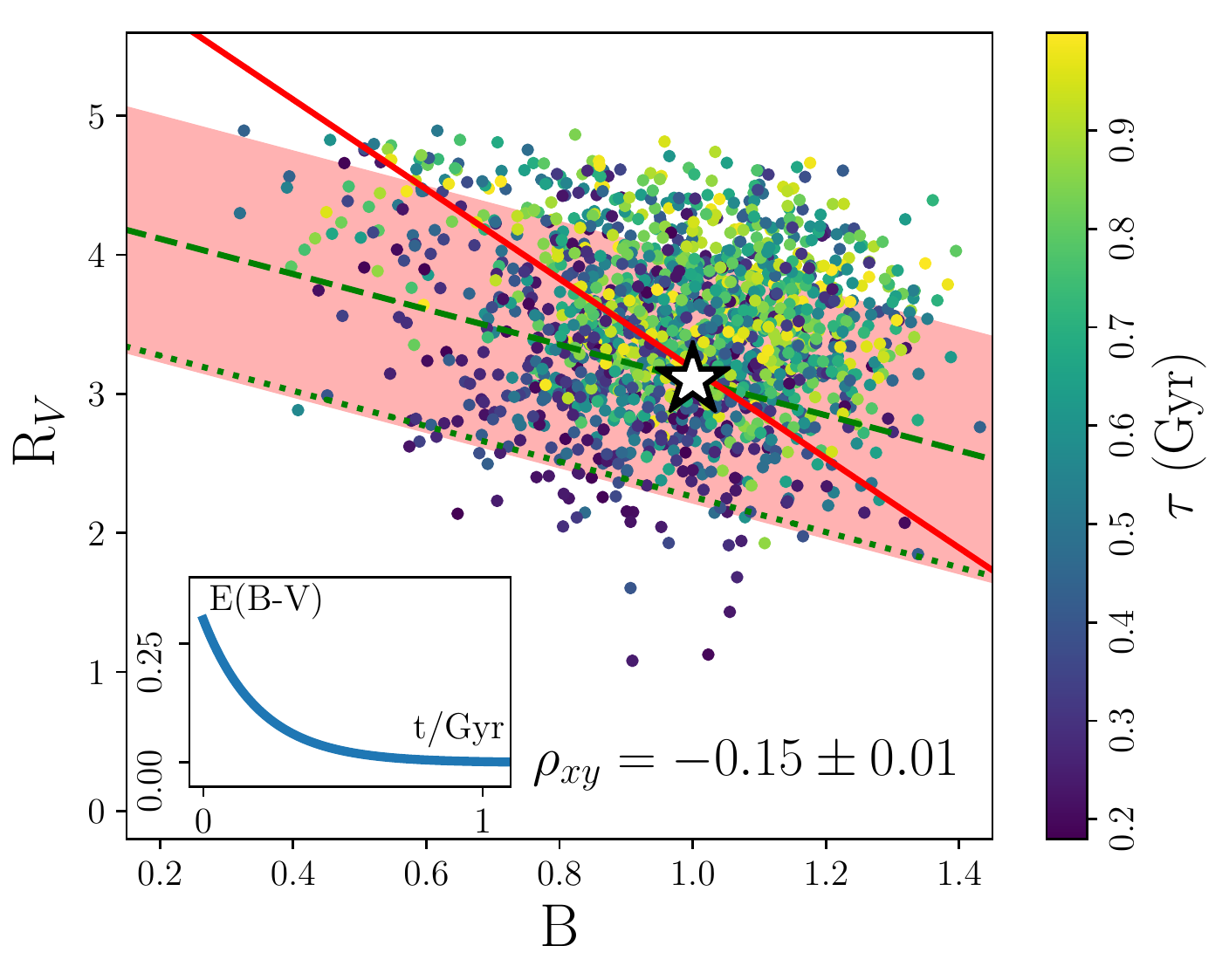}
\caption{Equivalent to Fig.~\ref{fig:mu} for a model where the
colour excess changes exponentially with time, given by a free
parameter $\tau$ (equation~\ref{eq:taumodel}, see an example in the
inset).  The star symbol marks the standard extinction law of the
Milky Way.  The dots are colour coded according to the value of $\tau$
(colour bar on the right). The red solid line shows the best-fit.
The linear correlation coefficient ($\rho_{xy}$) is included along
with a 1\,$\sigma$ error bar. The dashed line and shaded area show
the observed slope and scatter from \citet{TMF:18} when imposing the
Milky Way constraint at $B=1$, and the dotted line is the original
relation.}
\label{fig:tau}
\end{figure}

An additional test of the age-dependent extinction scenario is
to assess how the functional form of this dependence affects the derived
effective attenuation. We decided to explore an exponentially decaying
law, where the colour excess is described by:
\begin{equation}
E(B-V)(t) = \epsilon_M e^{-t/\tau},
\label{eq:taumodel}
\end{equation}
and $\tau$ introduces a timescale, representing the efficiency of
disintegration/dispersal of dust. The rate of change with time is
slowed in this model, with respect to the linear model with (log) time, for
the range of parameters explored. Note as $\tau\rightarrow 0$, we
concentrate the dust only on the very youngest phases of star formation,
whereas $\tau\rightarrow\infty$ represents a foreground dust screen
(as all populations are equally affected by the same level of extinction). To create
the mock data we follow a similar methodology as in the linear case
regarding the distribution of redshift, flux uncertainty and passband
offsets, as well as the underlying (constant) star formation history
between 0.1 and 1\,Gyr. This model has two free parameters: for each
galaxy a random pair $(\tau, \epsilon_M)$ is chosen from a uniform
random deviate, between $\tau$=0.1 and 1\,Gyr and between $\epsilon_M$=0.1 and
0.6\,mag. The synthetic data are analysed in the
same way as the observed SHARDS star-forming galaxies
of \citet{TMF:18}. Fig.~\ref{fig:tau} shows the trend of the
dust parameters on the $R_V$ vs $B$ plane.

A similar correlation is found between $R_V$ and bump strength (blue
solid line) as in the previous case, in qualitative agreement with the
observed trend, given by a dotted line. Analogously to the previous
case, we shift the trend by a constant value, using the Milky Way
constraint at $B=1$, represented by the dashed line and the shaded
area. The data suggest higher values of $\tau$ (i.e. a weaker age-dependent
extinction) produce a 
greyer attenuation law. The linear correlation coefficient is
$\rho_{xy}= -0.15\pm 0.01$. Therefore, at face value, this simple
exercise suggests that an age-dependent extinction provides a suitable
description of the destruction and dispersal of dust in star-forming
galaxies, and can explain -- at least in part -- the observed
anticorrelation between $R_V$ and bump strength.

\section{Comparison with standard dust attenuation models}
\label{Sec:Modelseq}

In addition to the phenomenological age-dependent extinction model shown
above, we compared the results of our SHARDS star-forming galaxies with two
of the standard dust attenuation models, \citet[]{DL:07}
and \citet{WG:00}.

\subsection{Draine \& Li models}
\label{Sec:Draineandli}

The models of \citet{DL:07} provide a way to derive fundamental
properties of the dust component from infrared observations with
Spitzer's IRAC and MIPS cameras. These models consist of a mixture of
Polyaromatic Hydrocarbons (PAHs), carbonaceous grains, and amorphous
silicate grains, heated by starlight. The main parameters are the mass
fraction of the dust content in the form of PAHs ($q_\mathrm{PAH}$,
usually given as a percentage), the lower cutoff of the starlight
intensity distribution ($U_\mathrm{min}$), the fraction of dust mass ($\gamma$) that is
exposed to starlight intensities $U_\mathrm{min}<U\leq U_\mathrm{max}$,
and the total dust mass ($M_d$).  The full set of SHARDS fluxes,
combined with {\sl Spitzer} and {\sl Herschel} data
\citep{SHARDS} was used to derive these parameters following the
prescriptions set out in \citet{DL:07}. The parameters  are 
fitted to galaxies with 24\,$\mu$m fluxes plus, at least, one flux
data point from {\sl Herschel}, needed to lift the degeneracy between the 
model parameters. Only galaxies with IR fluxes having S/N$\geq$3 are
considered in this analysis. Milky Way type dust was assumed. Given the very discrete
values of the dust model parameters published by \citet{DL:07}, we interpolated
them to obtain more models with a finer parameter resolution in
$\gamma$ and $q_\mathrm{PAH}$. 
The fitting procedure compares the data from the model grid to the
observed fluxes for each galaxy, following a
standard $\chi^2$ statistic, so that the best-fit parameters
correspond to the minimum value of the statistic. A Monte Carlo
method is adopted, where the fitting procedure is repeated 10 times
per galaxy, adding in each realization a Gaussian random deviate
to the observed fluxes with zero mean and variance equivalent to the
flux uncertainties. The distribution of best fit values for
each galaxy allows us to derive the uncertainty of the parameters.
We note that these
parameters are constrained with infrared fluxes, therefore requiring mostly
independent information with respect to the dust parameters presented
in \citet{TMF:18}, that focus on the rest-frame NUV and optical
windows. There is a total of 71 galaxies with good fits to these
models.  We note that the redshift distribution of this subsample is
undistinguisable from the original dataset, but the sample is biased
towards the massive end. The median and standard deviation of the stellar mass
is $\log M_s/M_\odot=10.53\pm 0.55$ in contrast with $9.59\pm 0.56$ for the
original sample.

\begin{figure*}
\centering
  \includegraphics[width=155mm]{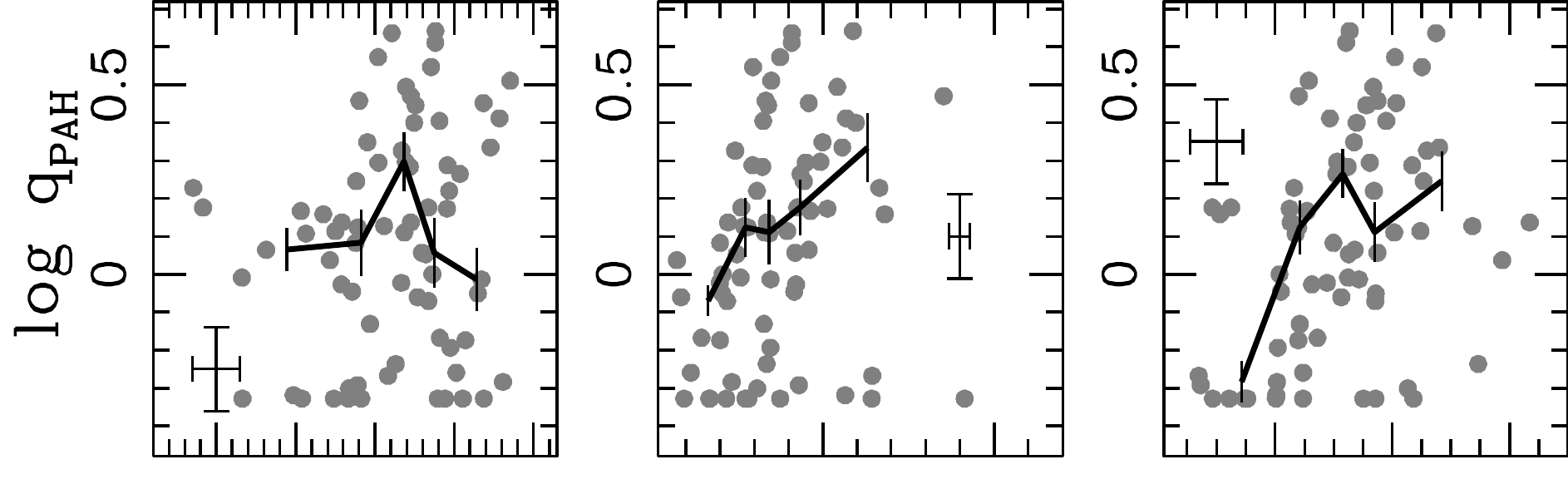}
  \includegraphics[width=155mm]{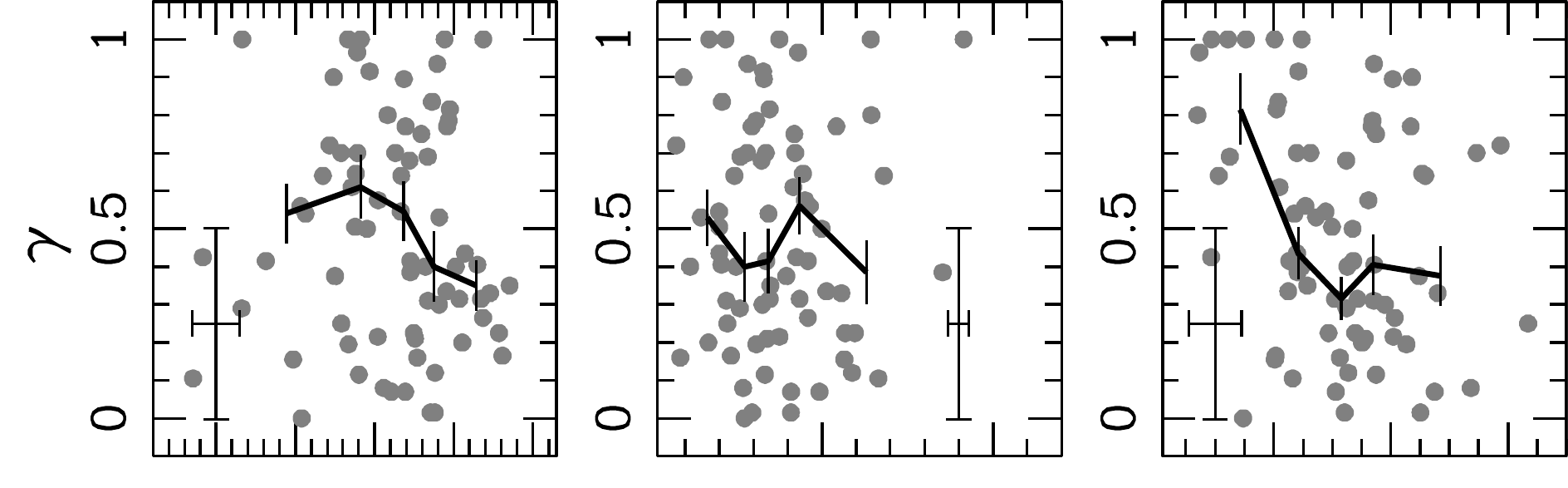}
  \includegraphics[width=155mm]{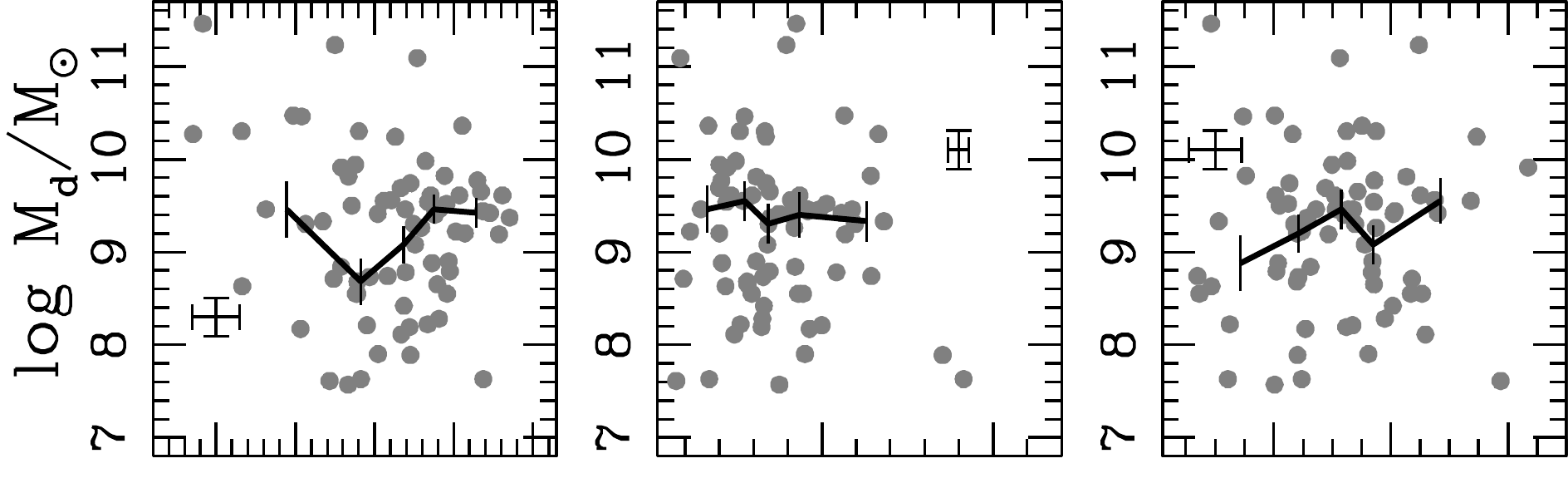}
  \includegraphics[width=155mm]{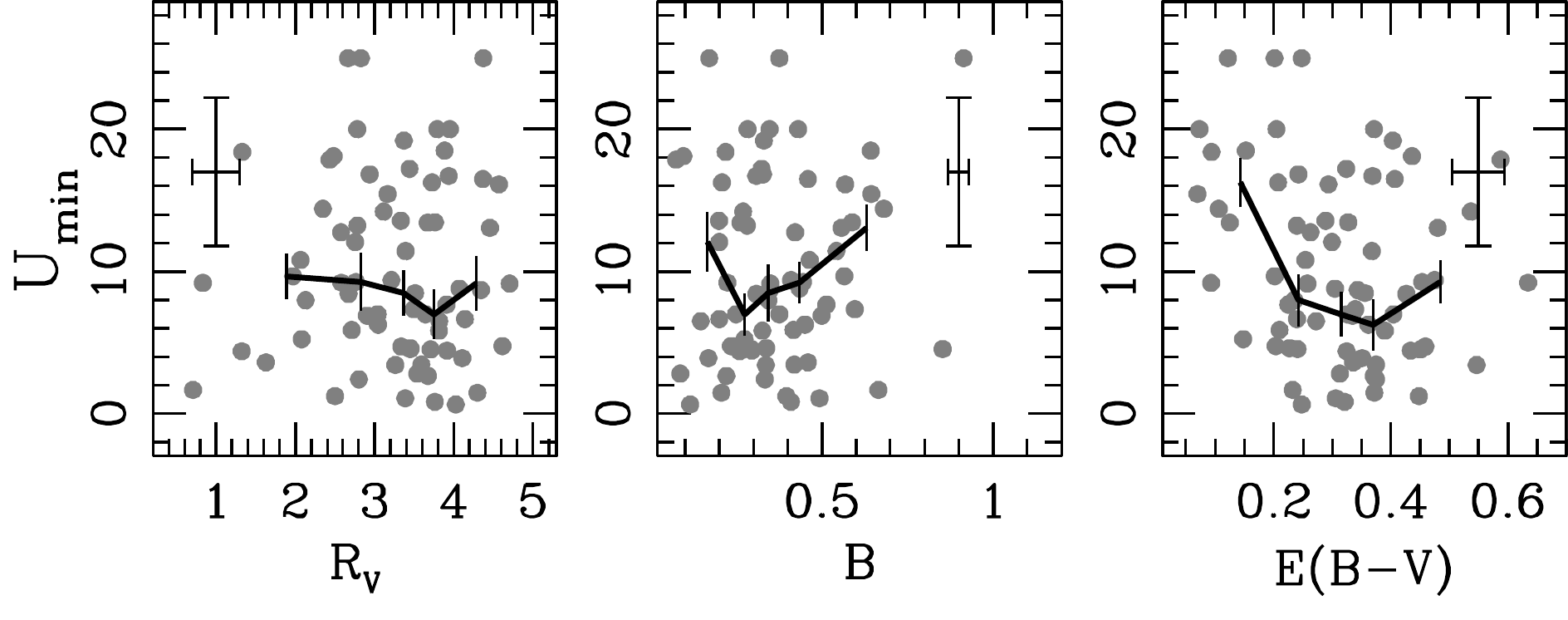}

  \caption{Distribution of dust parameters following the
    models of \citet{DL:07}, from top to bottom the PAH strength
    ($\log q_\mathrm{PAH}$), the fraction of dust mass ($\gamma$) that
    is exposed to starlight intensities $U_\mathrm{min}<U\leq U_\mathrm{max}$,
    the dust mass ($\log M_d$, in solar units),
    and the radiation field threshold ($U_{\rm min}$). A running median,
    including vertical lines to represent the RMS scatter within each bin,
    is shown as a solid line. The error bar in each panel gives a
    representative (median) 1\,$\sigma$ uncertainty of the parameter
    estimates, from a Monte Carlo run (see text for details).
}
\label{fig:DraineLi}
\end{figure*}

Figure \ref{fig:DraineLi} shows the constraints to the Draine \& Li
model parameters with respect to our dust-related parameters (from
left to right): $R_V$, $B$ and $E(B-V)$.  Individual data points are
shown as grey dots, whereas the solid lines follow the median of
binned data at a fixed number of data points per bin. The vertical bars
in the lines represent the median uncertainty at a 1\,$\sigma$ level.
A characteristic error bar is included in each panel, corresponding
to the median value of the 1\,$\sigma$ uncertainty of the relevant parameters.
The PAH fraction ($q_\mathrm{PAH}$) increases with $B$ and
colour excess, suggesting that dustier environments with
a prominent NUV bump feature a higher fraction of such molecules.
It is commonly accepted that graphite-like, very small dust particles
are the carrier of the NUV bump \citep[e.g.][]{Draine:89}.
The fraction of dust exposed to starlight ($\gamma$), within the
parameterised range of intensity, is found to
decrease with increasing colour excess, and possibly (although weakly so)
R$_V$, potentially as a consequence of shielding in dustier environments.
The trends of dust mass are rather weak, with prominent scatter,
and a hint of an increasing trend with colour excess. 
Finally, the starlight intensity distribution $U_\mathrm{min}$ is found
to decrease with increasing $E(B-V)$, and also to decrease
as the NUV bump becomes less prominent. However, we emphasize
that these trends should be considered semi-quantitative, as the range
of variation and the underlying uncertainties are rather large. More
detailed modelling would be necessary, but is beyond the scope of this
paper.

\subsection{Witt \& Gordon dust models}
\label{Ssec:WG}

One of the standard benchmarks to characterize the attenuation law in
galaxies is provided by the dust models of \citet[][hereafter
WG00]{WG:00}. The models explore three different galactic
environments: CLOUDY, DUSTY and SHELL. Each one features a different
distribution of dust with respect to the underlying stellar component.
Specifically, SHELL corresponds to a stellar region surrounded by a
dust shell; DUSTY has a uniform mixture of stars and dust and CLOUDY
is defined by a mixed dust-stellar region surrounded by a dust-free
region. In addition to the global dust geometry, these models also
adopt two different types of local structure: homogeneous and
clumpy. There are significant variations between these two.
For instance, the NUV bump strength is
weaker in the clumpy case, since it has a greyer wavelength dependence
of the attenuation, with respect to the homogeneous
structure \citep{WG:00}. WG00 models adopt two different intrinsic
extinction prescriptions, corresponding to the Milky Way and the bar
of the Small Magellanic Cloud.  The variations among these diverse
scenarios concern the different return of scattered radiation, producing a
range of effective attenuation laws. For instance, the homogeneous
SHELL model is comparable to a spherical homogeneous screen with flux
scattered back into the beam \citep{WG:00}.

We compare the functional dependence of the CSB10 dust attenuation
function (i.e. the $B$ and $R_V$ parameterisation) with respect to the
WG00 models. Moreover, we include the equivalent parameters $E_b$ and
$\delta$ in the notation of \citet{Noll:09}. This alternative
parameterisation defines the NUV bump as a Lorentzian-like Drude
profile (similarly but not equivalently to CSB10, so that parameters
$B$ are $E_b$ are related, but are not identical), and describes the
varying total-to-selective ratio as a multiplicative power law
$\propto\lambda^\delta$ applied to a Calzetti-like dust law (so that
$\delta=0$ corresponds to R$_V$=4.05, see \citealt{Noll:09} for
details). A translation between these two choices of the attenuation
law can be found in Appendix~A of \citet{TMF:18}.

We consider three
optical depths, namely $\tau_{V} = \{0.5, 1.5, 3.0\}$.  The comparison --
within the spectral window  $\lambda\in[1300,7500]$\,\AA -- 
is done with a {\sc Python} script that fits the publicly available WG00
attenuation curves with these models, following a two stage process,
starting with a Sequential Least SQuares
Programming\footnote{calling SciPy function {\tt optimize.minimize}
with method SLSQP.}  fit with a large error bar (assuming 20\% of the
value of the opacity at each wavelength), followed by an MCMC search
starting with the best-fit value using the {\sc emcee}
sampler \citep{emcee}, reducing in this second stage the error bars to
a 5\% level. The output after 400 steps (removing the first 100
burn-in steps) produces a distribution from which the median and RMS of each parameter
is obtained, shown in Tab.~\ref{table:WG}. The final result -- taking the
median of the distribution for each parameter as best fit -- is
plotted and visually inspected to confirm the convergence of the
fitting procedure. In general, a slightly steeper dust attenuation
curve is found in the homogeneous case, in comparison to the clumpy
scenario.  As the optical depth increases, the NUV bump strength
decreases for both CLOUDY and DUSTY geometries. This behaviour is
observed both in the homogeneous and clumpy cases.  In particular, in
the CLOUDY scenario, shifting from a moderate optical depth to a very
optically thick case results on the strength of the NUV bump to be
halved. The same occurs in a clumpy SHELL geometry. This is a
trend observed in galaxies at low redshift, where the NUV bump is
stronger at lower optical opacities \citep{SBL:18}.  On the other
hand, for the homogeneous case there is no significant change
regarding $B$.  In contrast, the total-to-selective ratio $R_{V}$
increases in every case, although for the SHELL homogeneous scenario,
this increase is less obvious.  As the optical depth increases, the
attenuation becomes greyer, a trend reported in several
samples \citep{SPL:16,TMF:18, SBL:18, NCDJP:18}.

\begin{figure}
\centering
\includegraphics[width=83mm]{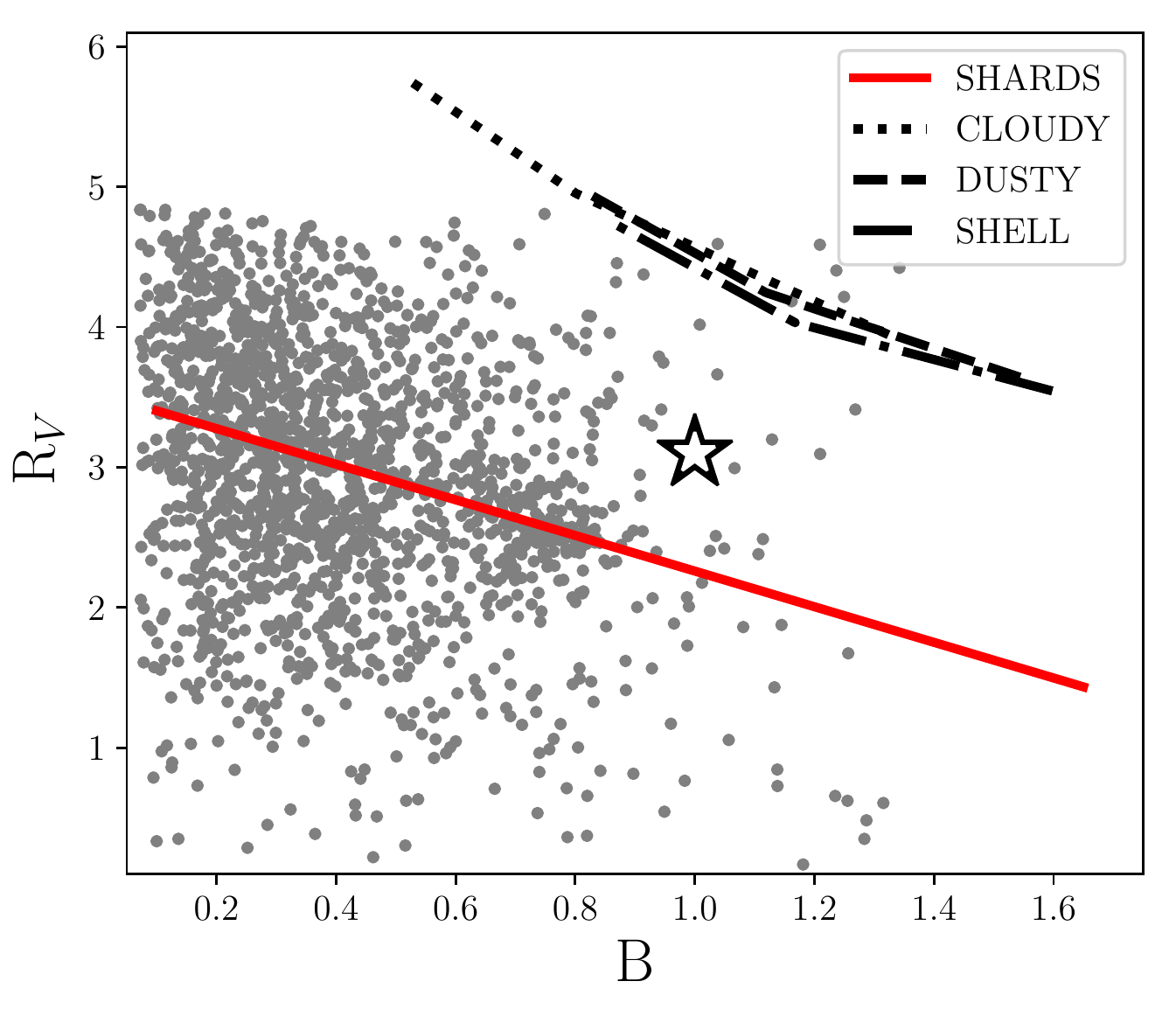}
\hspace*{-2.5mm}\includegraphics[width=85mm]{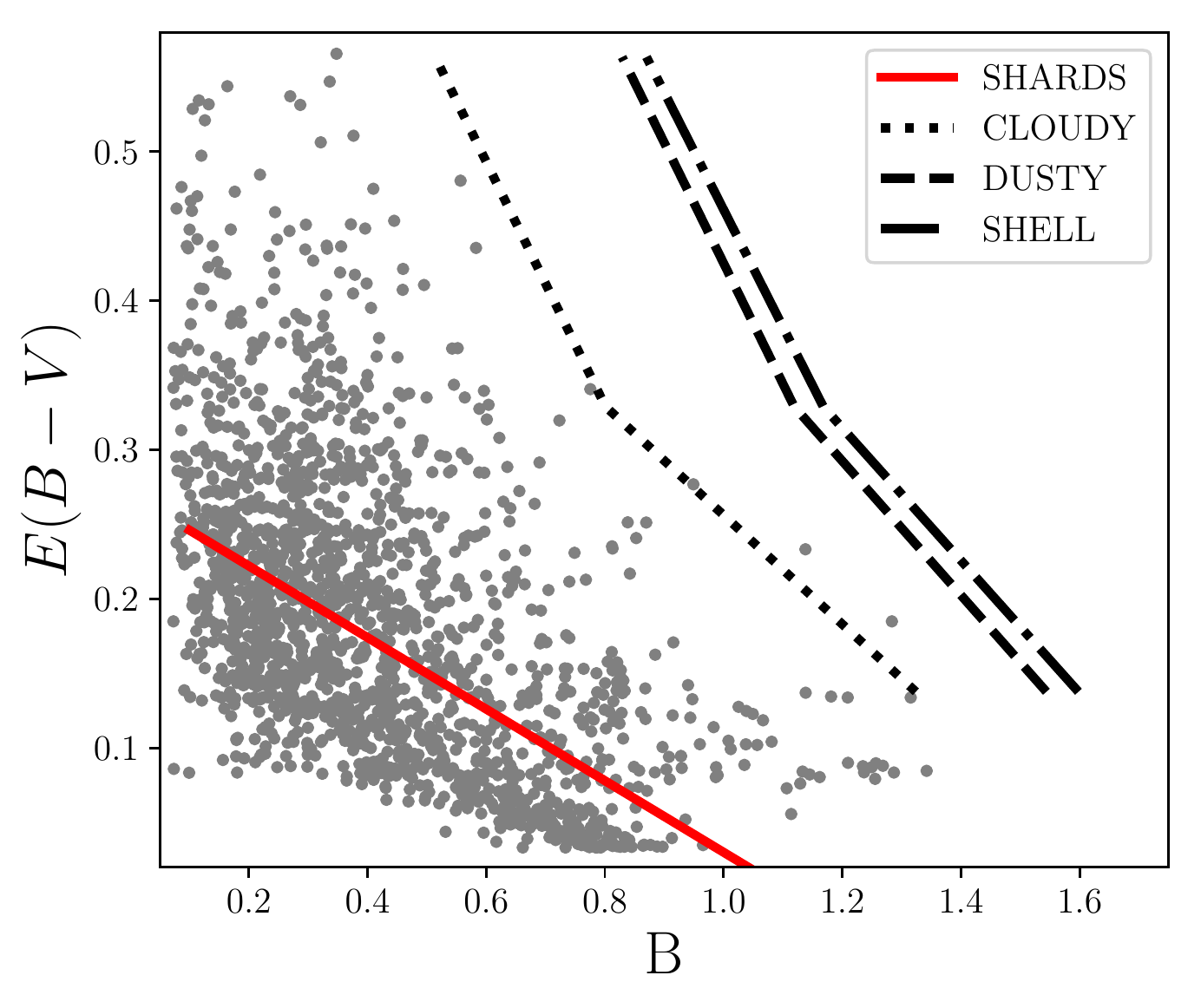}
\caption{Comparison of the \citet[][WG00]{WG:00} models 
to the SHARDS star-forming galaxy sample of \citet{TMF:18}
on the $R_V$ vs $B$ (top) and $E(B-V)$ vs $B$ planes (bottom).
Individual measurements are shown as grey dots, and the
best fit to the data is given by the solid red line.
The black lines, as labelled, correspond to the 
WG00 galactic environments, adopting a local clumpy
structure in all three cases. For reference, the
standard extinction law of the Milky Way is represented
by a star in the top panel.
}
\label{fig:WG00}
\end{figure}

Fig.~\ref{fig:WG00} compares the clumpy version of the WG00 models to
the observational constraints to the attenuation law of the SHARDS
star-forming galaxies from \citet{TMF:18}, showing the individual
galaxies as grey dots and the solid red line tracing the best fit to
the models. The star locates the fiducial Milky Way extinction law.
The WG00 models do not provide a good match to the data, mainly as the
best-fit values of $R_V$ are too high, producing substantially greyer
models, and also rather prominent NUV bump strengths. However, the trends have
similar slopes, from a steeper law with a strong bump (at low opacity)
towards a weaker bump with a greyer attenuation.  We note that the
WG00 models explored here adopt the Milky Way extinction law. We did
not fit the models with an SMC-type extinction, since, by definition,
the CSB10 function is parametrised taking the Milky Way extinction law
as a reference. Moreover, the lack of a bump in the extinction law of
the SMC would result in a trivial $B=0$ parameter in all cases.

\begin{table*}
\centering
\caption{Fit of the \citet{WG:00} models (Milky Way extinction) to the
generic attenuation function of \citet{CSB:10}.}
\label{table:WG}
\begin{tabular}{c|cccc|cccc}
\hline
$\tau_{V}$ & $B$ & $R_V$ & $E_b$ & $\delta$ & $B$ & $R_V$ & $E_b$ & $\delta$ \\
\hline
\multicolumn{9}{c}{CLOUDY}\\
\hline
   & \multicolumn{4}{c}{homogeneous} &  \multicolumn{4}{c}{clumpy}\\
\hline
0.5 & $1.46\pm 0.10$ & $3.77\pm 0.07$ & $5.89\pm 0.46$ & $+0.13\pm 0.01$ & $1.32\pm 0.10$ & $3.95\pm 0.07$ & $5.04\pm 0.43$ & $+0.17\pm 0.01$\\
1.5 & $0.91\pm 0.09$ & $4.61\pm 0.08$ & $2.77\pm 0.37$ & $+0.31\pm 0.01$ & $0.80\pm 0.09$ & $4.95\pm 0.09$ & $2.24\pm 0.35$ & $+0.38\pm 0.01$\\
3.0 & $0.59\pm 0.09$ & $5.53\pm 0.09$ & $1.37\pm 0.33$ & $+0.49\pm 0.01$ & $0.51\pm 0.09$ & $5.78\pm 0.10$ & $1.14\pm 0.32$ & $+0.54\pm 0.01$\\
\hline
\multicolumn{9}{c}{DUSTY}\\
\hline
   & \multicolumn{4}{c}{homogeneous} &  \multicolumn{4}{c}{clumpy}\\
\hline
0.5 & $1.69\pm 0.11$ & $3.50\pm 0.06$ & $7.38\pm 0.51$ & $+0.06\pm 0.01$ & $1.54\pm 0.11$ & $3.64\pm 0.06$ & $6.46\pm 0.48$ & $+0.10\pm 0.01$\\
1.5 & $1.33\pm 0.10$ & $3.85\pm 0.07$ & $5.16\pm 0.44$ & $+0.15\pm 0.01$ & $1.12\pm 0.10$ & $4.25\pm 0.07$ & $3.87\pm 0.40$ & $+0.24\pm 0.01$\\
3.0 & $1.08\pm 0.10$ & $4.38\pm 0.07$ & $3.62\pm 0.40$ & $+0.26\pm 0.01$ & $0.83\pm 0.09$ & $4.94\pm 0.08$ & $2.34\pm 0.36$ & $+0.38\pm 0.01$\\
\hline
\multicolumn{9}{c}{SHELL}\\
\hline
   & \multicolumn{4}{c}{homogeneous} &  \multicolumn{4}{c}{clumpy}\\
\hline
0.5 & $1.84\pm 0.11$ & $3.28\pm 0.06$ & $8.61\pm 0.53$ & $+0.01\pm 0.01$ & $1.60\pm 0.11$ & $3.54\pm 0.06$ & $6.87\pm 0.49$ & $+0.08\pm 0.01$\\
1.5 & $1.74\pm 0.10$ & $3.22\pm 0.05$ & $8.17\pm 0.52$ & $-0.00\pm 0.01$ & $1.17\pm 0.10$ & $4.03\pm 0.07$ & $4.27\pm 0.41$ & $+0.19\pm 0.01$\\
3.0 & $1.65\pm 0.11$ & $3.23\pm 0.05$ & $7.69\pm 0.52$ & $-0.00\pm 0.01$ & $0.87\pm 0.09$ & $4.73\pm 0.08$ & $2.58\pm 0.37$ & $+0.33\pm 0.01$\\
\hline
\end{tabular}
\end{table*}

\section{Discussion}
\label{Sec:Disc}

This paper extends the analysis of the sample of high
redshift, star-forming galaxies presented in \citet{TMF:18}. In that
paper, we exploited the SHARDS survey
\citep{SHARDS} to
constrain the dust attenuation law of star-forming galaxies over the
redshift interval 1.5$<$z$<$3. A substantial correlation was found
between the total-to-selective extinction (parameterised by $R_{V}$), 
the strength of the NUV bump, and the colour excess. The interpretation
involves two main aspects of dust in galaxies: its intrinsic
composition (affecting the extinction law) and the distribution of
dust within the underlying stellar components (the so-called
``geometry''). We explore in this paper an eclectic set of potential
ramifications of the wide range of dust attenuation parameters found
in star-forming galaxies, mostly related to diagnostics of star formation
vs quiescence or to the interpretation of the observed variations in
light of a few well established models of dust attenuation.

Our first aim concerns the characterisation of the sample of SHARDS
galaxies regarding star formation activity, by putting them on a rest-frame UVJ
colour-colour diagram. Such diagrams are typically used to classify galaxies as either 
quiescent or star-forming systems. The wide range of dust attenuation
parameters found in \citet{TMF:18} raised the issue whether the standard classification
based on this bicolour plot could
be affected by it. Our analysis showed that the loci used to define
quiescent galaxies was not compromised as the observed variations
followed the same direction on the colour-colour diagram as the
overall extinction $A_V$ (Fig.~\ref{fig:uvjdust}).  However, these
variations introduce a more complex degeneracy along the diagonal
direction of the diagram with respect to age, metallicity, colour
excess {\sl and} total-to-selective extinction ($R_V$). 
The strength of the NUV bump is also found to correlate with the
location of the galaxies on the UVJ plane (note that 
this feature is rather wide, so it can affect the fluxes through the $U$ passband).
We also assessed a potential correlation of the dust-parameters with
respect to the location of these galaxies relative to the main sequence of
star formation. Most of our sample falls close to the main sequence,
with variations within the 25\% level (Fig.~\ref{fig:MShist}), with no
statistically significant correlation with the dust parameters (Fig.~\ref{fig:MSdust}).

We explored the efficiency of dust destruction/dispersal with a simple
phenomenological model given by an age-dependent reddening. We used two different
efficiencies, a linear decrease of reddening with (log) time
(Fig.~\ref{fig:mu}); and an exponential decrease
(Fig.~\ref{fig:tau}). Both scenarios are extensions of the `birth
cloud' model over longer timescales, whereby younger stars suffer more dust attenuation compared
to older stars. The timescale of our phenomenological model includes
the contribution from additional mechanisms that regulate the global evolution of dust
over galaxy scales. We built mock data from the \citet{BC03}
population synthesis models, applying an age-dependent attenuation with a fixed
(Milky Way standard) extinction, as a test of possible causes of
the correlations found in \citet{TMF:18}. We find that this simple
model indeed produces an anticorrelation between $R_V$ and $B$, so that
galaxies with a weaker NUV bump feature greyer
attenuation, supporting earlier claims (see, e.g., \citealt{PGB:07};
\citealt{NPS:07}; \citealt{Buat:11}; \citealt{NCDJP:18}).

Finally, we compared the observational data of SHARDS star-forming galaxies with
the standard dust models of \citet[]{DL:07} and
\citet[]{WG:00}. Fig.~\ref{fig:DraineLi} shows 
the best fits to the \citet[]{DL:07} models with respect to $E(B-V)$,
$R_{V}$ and $B$.  The \citet[]{DL:07} parameters are the PAH fraction
($q_\mathrm{PAH}$); the fraction of dust mass ($\gamma$) that is
exposed to starlight intensities $U_\mathrm{min}<U\leq U_\mathrm{max}$;
the dust mass ($M_d$); and the lower cutoff of the
starlight intensity distribution ($U_\mathrm{min}$).
The increasing trend of $q_\mathrm{PAH}$ with $B$ and colour excess
(Fig.~\ref{fig:DraineLi}) is suggestive of an increased contribution from PAH
molecules in dustier environments with a more prominent bump strength
Although weak, the data also reveal a correlation between dust mass and $E(B-V)$.
Furthermore, we calculated the $R_{V}$ and NUV
bump strength equivalents of the \citet[]{WG:00} models. The results are
shown in Table~\ref{table:WG} and Fig.~\ref{fig:WG00}. As the optical depth
increases, the attenuation becomes greyer and the bump strength weaker
\citep[Fig.~\ref{fig:WG00}, see also][]{SPL:16,TMF:18, SBL:18, NCDJP:18}.
We note that the observed anti-correlation between $R_{V}$ and $B$,
i.e. greyer attenuation associated to a smaller bump, is also seen in
other studies \citep[][]{KC:13,SD:16, TMF:18, SBL:18, NCDJP:18}.
Two alternative options can be considered to explain these correlations,
invoking variations in either dust composition or dust geometry. Our
analysis -- based on simple extensions of the birth cloud model -- 
suggests that geometry is potentially the major driver of these
trends.


\label{lastpage}

\end{document}